\documentclass[letter, longauth]{aa}

\relax
\usepackage{graphicx}
\usepackage{txfonts}
\usepackage{pdflscape}
\usepackage{color}

\usepackage{amstext}
\usepackage[colorlinks=true,linkcolor=black,citecolor=blue]{hyperref}%

\begin{document} 

\title{The Galaxy Activity, Torus, and Outflow Survey (GATOS). III: Revealing the inner icy structure in local AGN}

   \author{I. Garc\'ia-Bernete\inst{1}\fnmsep\thanks{E-mail: igbernete@gmail.com}, A. Alonso-Herrero\inst{2}, D. Rigopoulou\inst{1,3}, M. Pereira-Santaella\inst{4}, T. Shimizu\inst{5}, R. Davies\inst{5}, F.\,R. Donnan\inst{1}, P.\,F. Roche\inst{1}, O. Gonz\'alez-Mart\'in\inst{6}, C. Ramos Almeida\inst{7,8}, E. Bellocchi\inst{9,10}, P. Boorman\inst{11}, F. Combes\inst{12}, A. Efstathiou\inst{3}, D. Esparza-Arredondo\inst{7,8}, S. Garc\'ia-Burillo\inst{13}, E. Gonz\'alez-Alfonso\inst{14}, E.\,K.\,S. Hicks\inst{15}, S. H\"onig\inst{16}, A. Labiano\inst{2,17}, N. A. Levenson\inst{18}, E. L\'opez-Rodr\'iguez\inst{19}, C. Ricci\inst{20}, C. Packham\inst{21}, D. Rouan\inst{12}, M. Stalevski\inst{22} and M.\,J.\,Ward\inst{23}.}

   \institute{$^1$Department of Physics, University of Oxford, Keble Road, Oxford OX1 3RH, UK \\
   $^2$Centro de Astrobiolog\'ia (CAB), CSIC-INTA, Camino Bajo del Castillo s/n, E-28692, Villanueva de la Ca\~nada, Madrid, Spain\\
   $^{3}$School of Sciences, European University Cyprus, Diogenes street, Engomi, 1516 Nicosia, Cyprus\\   
   $^4$Instituto de F\'isica Fundamental, CSIC, Calle Serrano 123, 28006 Madrid, Spain\\
   $^5$Max-Planck-Institut fur extraterrestrische Physik, Postfach 1312, D-85741 Garching, Germany\\
   $^6$Instituto de Radioastronom\'ia and Astrof\'isica (IRyA-UNAM), 3-72 (Xangari), 8701, Morelia, Mexico\\ 
   $^7$Instituto de Astrof\'isica de Canarias, Calle V\'ia L\'actea, s/n, E-38205 La Laguna, Tenerife, Spain\\
   $^8$Departamento de Astrof\'isica, Universidad de La Laguna, E-38206 La Laguna, Tenerife, Spain\\
   $^9$Departamento de F\'isica de la Tierra y Astrof\'isica, Fac. de CC F\'isicas, Universidad Complutense de Madrid, 28040 Madrid, Spain\\
   $^{10}$Instituto de F\'isica de Part\'iculas y del Cosmos IPARCOS, Fac. de CC F\'isicas, Universidad Complutense de Madrid, 28040 Madrid, Spain\\
   $^{11}$Cahill Center for Astrophysics, California Institute of Technology, 1216 East California Boulevard, Pasadena, CA 91125, USA\\
   $^{12}$Observatoire de Par\'is, PSL Research University, CNRS, Sorbonne Universit\'es, UPMC Univ. Paris 06, 92190 Meudon, France\\
   $^{13}$Observatorio Astron\'omico Nacional (OAN-IGN)-Observatorio de Madrid, Alfonso XII, 3, 28014, Madrid, Spain\\
   $^{14}$Universidad de Alcal\'a, Departamento de F\'isica y Matem\'aticas, Campus Universitario, 28871, Alcal\'a de Henares, Madrid, Spain\\
   $^{15}$Department of Physics \& Astronomy, University of Alaska Anchorage, AK 99508-4664, USA\\
   $^{16}$Department of Physics \& Astronomy, University of Southampton, Hampshire, SO17 1BJ, Southampton, UK\\
   $^{17}$Telespazio UK for the European Space Agency (ESA), ESAC, Camino Bajo del Castillo s/n, 28692 Villanueva de la Ca\~nada, Spain\\
   $^{18}$Space Telescope Science Institute, 3700 San Martin Drive, Baltimore, Maryland 21218, USA\\
   $^{19}$Kavli Institute for Particle Astrophysics and Cosmology (KIPAC), Stanford University, Stanford, CA 94305, USA\\
   $^{20}$Instituto de Estudios Astrof\'isicos, Facultad de Ingenier\'ia y Ciencias, Universidad Diego Portales, Avenida Ejercito Libertador 441, Santiago, Chile\\
   $^{21}$The University of Texas at San Antonio, One UTSA Circle, San Antonio, TX 78249, USA\\
   $^{22}$Astronomical Observatory, Volgina 7, 11060 Belgrade, Serbia\\
   $^{23}$Centre for Extragalactic Astronomy, Durham University, South Road, Durham DH1 3LE, UK\\ }

\titlerunning{Revealing the inner icy structure in local AGN}
\authorrunning{Garc\'ia-Bernete et al.}

   \date{}

  \abstract
   {We use JWST/MIRI MRS spectroscopy of a sample of six local obscured type 1.9/2 active galactic nuclei (AGN) to compare their nuclear mid-IR absorption bands with the level of nuclear obscuration traced by X-rays. This study is the first to use sub-arcsecond angular resolution data of local obscured AGN to investigate the nuclear mid-IR absorption bands with a wide wavelength coverage (4.9-28.1 $\mu$m). All the nuclei show the 9.7\,$\mu$m silicate band in absorption. We compare the strength of the 9.7 and 18\,$\mu$m silicate features with torus model predictions. The observed silicate features are generally well explained by clumpy and smooth torus models. We report the detection of the 6\,$\mu$m {\textit{dirty water ice}} band (i.e., a mix of water and other molecules such as CO and CO$_2$) at sub-arcsecond scales ($\sim$0.26\arcsec at 6\,$\mu$m; inner $\sim$50\,pc) in a sample of local AGN with different levels of nuclear obscuration in the range log N$_{\rm H}^{\rm X-Ray}$(cm$^{-2}$)$\sim22-25$. We find a good correlation between the 6\,$\mu$m water ice optical depths and N$_{\rm H}^{\rm X-Ray}$. This result indicates that the water ice absorption might be a reliable tracer of the nuclear intrinsic obscuration in AGN. The weak water ice absorption in less obscured AGN (log N$_H^{X-ray}$ (cm$^{-2}$)$\lesssim$23.0 cm$^{-2}$) might be related to the hotter dust temperature ($>$T$_{sub}^{H_2O}\sim$110\,K) expected to be reached in the outer layers of the torus due to their more inhomogeneous medium. Our results suggest it might be necessary to include the molecular content, such as, H$_2$O, aliphatic hydrocarbons (CH-) and more complex PAH molecules in torus models to better constrain key parameters such as the torus covering factor (i.e. nuclear obscuration).}

      
   \maketitle


\section{Introduction}

Active galactic nuclei (AGN) are considered to be a short-lived ($<$100~Myr; e.g. \citealt{Hopkins05}) but recurrent phase that might take place in all relatively massive galaxies (e.g. \citealt{Hickox14}). AGN are powered by accretion of material onto supermassive black holes (SMBHs), which releases energy in the form of radiation and/or mechanical outflows to the interstellar medium (ISM) of the host galaxy. 

Subarcsecond angular observations ($<$0.5\arcsec) of local AGN ($\sim$tens of Mpc) enable us to probe their nuclear/circumnuclear regions (inner $\sim$100~pc scales) where part of the present AGN feedback is taking place. At nuclear scales (tens of pc) the bulk of dust and gas surrounding the AGN was proposed to be distributed in a toroidal structure where the dust obscuration is significant (\citealt{Antonucci93}). ALMA observations detected the molecular dusty torus in several nearby AGN and showed that it is part of the galaxy gas flow cycle (e.g. \citealt{Garcia-Burillo16,Garcia-Burillo19,Garcia-Burillo21,Imanishi18,Imanishi20,Alonso-Herrero18,Alonso-Herrero19,Alonso-Herrero23,Combes19}). Depending on its orientation, it obscures the central engines of type 2 AGN, and while providing a direct view of the central engine in the case of type 1 AGN. This nuclear dust absorbs a significant part of the AGN radiation and then reprocesses it to emerge in the infrared (IR; e.g.  \citealt{Pier92}). Typically, due to the small angular size of this dusty structure, it is only resolved with interferometric observations (e.g. \citealt{Garcia-Burillo16,Garcia-Burillo21} at submm, \citealt{GRAVITY20} at near-IR, and \citealt{Rosas22} at mid-IR wavelengths) but not with single-dish 10-m class telescopes. Thus, an extensively used technique for constraining the nuclear dusty structure properties is to compare torus models to the observed nuclear IR emission (see \citealt{Ramos17} for a review). In particular, previous studies reported reasonably accurate fits to the nuclear near-IR to mid-IR SED of nearby AGNs using clumpy torus models (e.g., \citealt{Ramos09,Ramos11b,Hoenig10B,Alonso-Herrero11,Lira13,Ichikawa15,Garcia-Bernete15,Garcia-Bernete19,Garcia-Bernete22c,Martinez-Paredes15,Martinez-Paredes20,Martinez-Paredes21,Esparza-Arredondo19,Gonzalez-Martin19A,Gonzalez-Martin19B,Gonzalez-Martin23})

Silicate grains are an important component of the interstellar dust mix (standard values of the Milky way are: silicate $\sim$53\% and graphite 47\%; e.g. \citealt{Mathis77}). While graphites are features-less in the mid-IR, silicates produce absorption/emission bands at $\sim$9.7 and 18\,$\mu$m (e.g. \citealt{Ossenkopf92}). The mid-IR spectra of buried sources, like deeply embedded massive proto-stars, show strong absorption features produced by dust and icy material. In particular, H$_2$O ice absorption features are detected in these sources (at $\sim$3 and 6\,$\mu$m; see \citealt{Boogert15} for a review).  Water is abundant in the interstellar medium and the freezing of water vapour onto dust grain mantles is considered a vital mechanism (catalyst) to form molecules in the Universe (e.g. \citealt{Cazaux16}).

H$_2$O features are generally detected toward embedded young stellar objects. The first detection of 6\,$\mu$m water ice in extragalactic sources occurred in NGC\,4418 (\citealt{Spoon01}). To date, extra-galactic H$_2$O ices were detected in luminous infrared galaxies (log L$_{\rm IR}>$11 L$\sun$; e.g. \citealt{Spoon22} and references therein), which are extremely rich in molecular gas and dust. Given that water-ice and aliphatic hydrocarbon (CH-) absorption mainly occurs in sources with signatures of a buried nuclei \citep{Spoon22}. 

However, the unprecedented combination of high angular and spectral resolution (R$\sim$1500-3500) in the entire mid-IR range (4.9-28.1 $\mu$m) afforded  by the 6.5m \emph{James Webb} Space Telescope (JWST; \citealt{Gardner23})/Mid-Infrared Instrument (MIRI; \citealt{Rieke15, Wells15, Wright15, Wright23}) allows for detailed studies of the central and circumnuclear regions of local galaxies and allows searching for weak and/diluted features. This letter reports the detection of the 6\,$\mu$m water ice feature at sub-arcsecond scales ($\sim$0.26\arcsec at 6\,$\mu$m; in the inner $\sim$50\,pc) in a sample of local obscured AGN. Using JWST/MRS data, we find that water ices may be a reliable tracer of the nuclear intrinsic obscuration in AGN. The luminosity distance and spatial scale were calculated using a cosmology with H$_0$=70 km~s$^{-1}$~Mpc$^{-1}$, $\Omega_m$=0.3, and $\Omega_{\Lambda}$=0.7.

\begin{figure*}
\centering
\par{
\includegraphics[width=16.3cm]{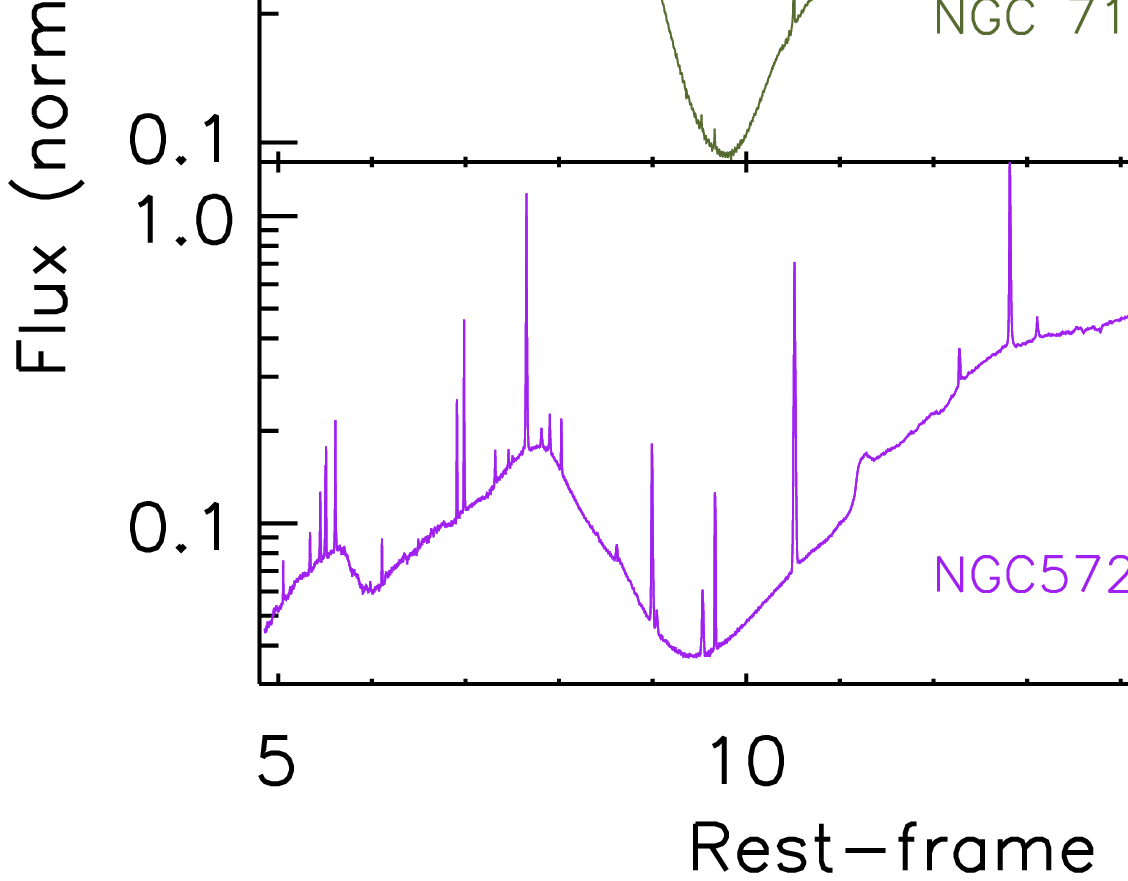}
\par}
\caption{JWST/MRS nuclear spectra of our objects drawn from the GATOS sample.} 
\label{gatos}
\end{figure*}

\begin{figure*}
\centering
\par{
\includegraphics[width=16.3cm]{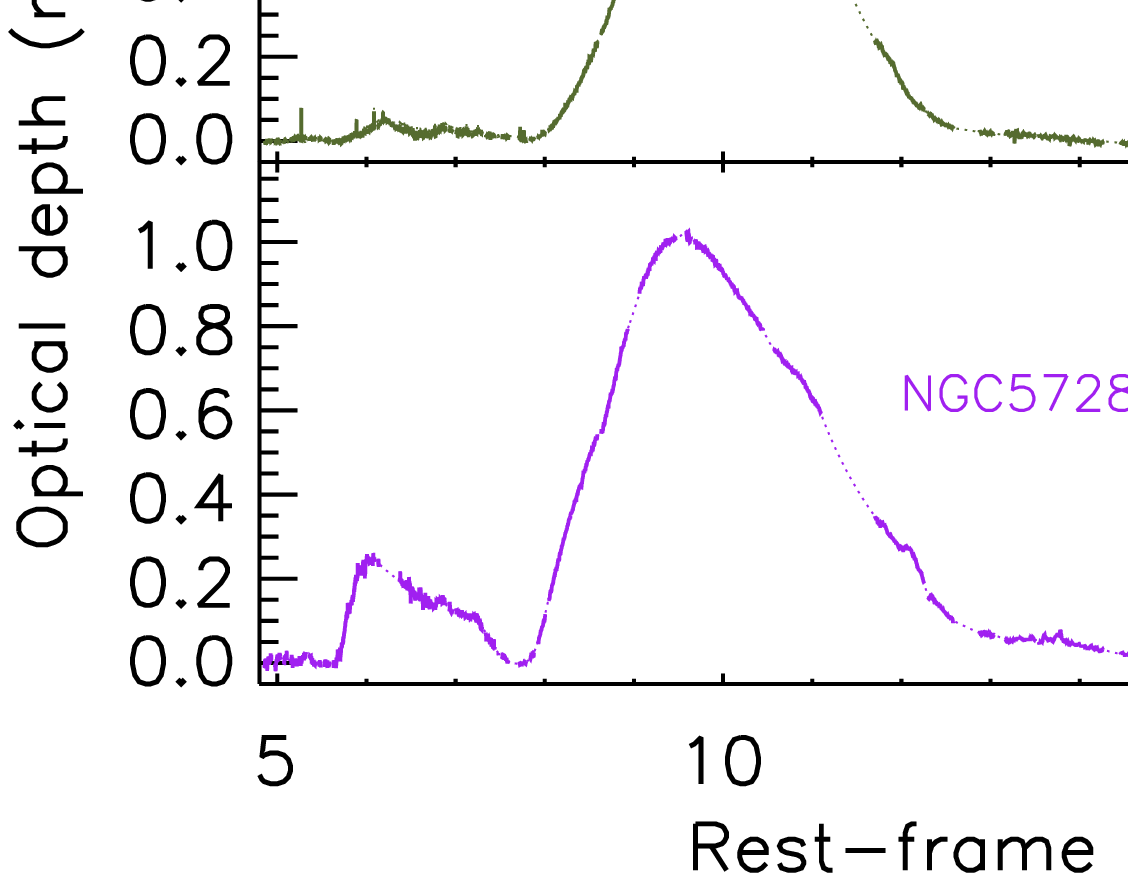}
\par}
\caption{Optical depth of the sample. All curves are normalized at 9.7\,$\mu$m. Dotted lines correspond to the masked spectral regions for removing the contribution of PAH features and narrow emission lines.}
\label{gatos_opticaldepth}
\end{figure*}
\section{Targets and observations}
\label{sample}
The galaxies studied here are part of the Galactic Activity, Torus, and Outflow Survey (GATOS \citealt{Garcia-Burillo21,Herrero21})\footnote{\textcolor{blue}{https://gatos.myportfolio.com/}}, which has the main goal of understanding the properties of the dusty molecular torus and its connection to their host galaxies in local AGN. The parent sample is selected from the 70th Month {\textit{Swift}}/BAT AGN catalog, which is flux-limited in the ultra-hard 14--195~keV X-rays band (\citealt{Baumgartner13}). The present study employs MIRI/MRS observations of a sub-sample of obscured type 1.9/2 AGN from the GATOS sample. These data are part of the JWST cycle 1 GO proposal ID\,1670 (PI: T. Shimizu and R. Davies). All the galaxies are classified as obscured AGN (log N$_H^{X-ray}$ (cm$^{-2}$)$>$22.0), and two of them are also classified as Compton thick (CT) AGN (NGC\,5728 and ESO\,137--G034; i.e. log N$_H^{X-ray}$ (cm$^{-2}$)$>$24.0; e.g. \citealt{Ricci17}). We also include a slightly more distant galaxy, NGC\,7319 (D$\sim$100\,Mpc), as it is an obscured type AGN, which was observed as part of the Early Release Observations (Program ID 2732, PI K. M. Pontoppidan; \citealt{Pontoppidan22}), which is publicly available in the JWST archive (these data were already presented in \citealt{Pereira22,Bernete22d}). The main properties of the sample are summarized in Table\,\ref{table_prop}. 

The sample was observed with MIRI/MRS (integral-field spectroscopy), which comprises four wavelength channels: ch1 (4.9--7.65~$\mu$m), ch2 (7.51--11.71~$\mu$m), ch3 (11.55--18.02~$\mu$m), and ch4 (17.71--28.1~$\mu$m). We refer the reader to Appendix \ref{reduction} for further details on the observations and data reduction. The newly observed JWST/MRS galaxies in Cycle\,1 are presented in Fig. \ref{gatos}. For comparison with the absorption bands in our sample, we use the Spitzer/IRS data of the deeply obscured type 2 NGC\,4418 (see Appendix \ref{ngc4418}).

To extract the JWST/MRS spectra from the nuclear regions (see Fig. \ref{gatos}), we used a point source extraction. See Appendix \ref{reduction} for further details on the spectral extraction. We note that the nuclear spectra of MCG-05-23-016, NGC\,5506, NGC\,5728, NGC\,7172, NGC\,7319, and ESO\,137-G034 correspond to physical scales of $\sim$57, 55, 53, 38, 55, 116 and 52\,pc (at 6\,$\mu$m), respectively.

\begin{table*}[ht]
\centering
\begin{tabular}{llcccccc}
\hline
Name & AGN     &D$_{\rm L}$ & $\tau_{6.0\mu m}$& $\tau_{9.7\mu m}$& $\tau_{18\mu m}$& log N$_{\rm H}^{\rm X-ray}$\\
 & type&    (Mpc) & &&&(cm$^{-2}$)\\
\hline
MCG-05-23-016& 1.9 & 35 &$<$0.01& 0.32$\pm$0.01& 0.01$\pm$0.01&22.2\\
NGC\,5506 &1.9 & 27  &0.08$\pm$0.02& 0.90$\pm$0.04& 0.10$\pm$0.02&22.4\\
NGC\,7172& 2& 37 &0.07$\pm$0.01& 2.24$\pm$0.03&  0.54$\pm$0.05&22.9\\
NGC\,7319& 2& 96 &0.05$\pm$0.02& 0.59$\pm$0.02&  0.01$\pm$0.01&23.8\\
NGC\,5728& 1.9& 39 &0.47$\pm$0.02& 1.91$\pm$0.02& 0.23$\pm$0.04&24.2\\
ESO\,137-G034& 2 & 35 &0.21$\pm$0.07& 0.99$\pm$0.03& 0.05$\pm$0.04&24.3\\
NGC\,4418&2 & 37 & 0.93$\pm$0.03& 4.10$\pm$0.01&0.95$\pm$0.01& $>$25\\
\hline
\end{tabular}                                           
\caption{Properties of the local AGN used in this work sorted by N$_{\rm H}^{\rm X-ray}$. The hydrogen column densities are from \citet{Sakamoto13} and \citet{Ricci17}. Note that for NGC\,4418 we are using Spitzer/IRS data. The spectral types are from \citet{Veron06}. $\tau_{6.0\mu m}$, $\tau_{9.7\mu m}$ and $\tau_{18.0\mu m}$ correspond the measured optical depth at 6.0, 9.7 and 18.0\,$\mu$m (see Section \ref{silicates} for details), respectively.}
\label{table_prop}
\end{table*}

\section{Nuclear absorption features}
\label{tauice}

\subsection{Silicate bands}
\label{silicates}
The silicate strength is sensitive to the balance of cold (absorption) and hot (emission) dust from various regions. This is especially relevant in clumpy distributions where the silicate band is filled by the emission from hotter clumps, resulting in weaker silicate absorption in the outer parts of the torus (e.g. \citealt{Levenson07,Nenkova08A,Nenkova08B,Nikutta21}). Furthermore, using subarcsecond resolution data with 8-10\,m ground-based telescopes, it has been found that foreground obscuration by the host galaxy might be important at these scales (e.g. \citealt{Alonso-Herrero11, Gonzalez-Martin13, Garcia-Bernete19,Garcia-Bernete22c}). At nuclear scales (few tens of pc), silicate emission is not always observed in type 1 AGN which can also show weak absorption and type 2 AGN generally have moderate silicate absorption features (see e.g. \citealt{Roche07,Alonso-Herrero11,Alonso-Herrero16,Hoenig10B,Gonzalez-Martin13,Garcia-Bernete17,Garcia-Bernete19,Garcia-Bernete22c}). 

We measure the silicate strengths at 9.7 and 18\,$\mu$m in the nuclear spectra for all the AGN in the sample. We computed the silicate strength (S$_{\rm Sil}^{\lambda}$) following a similar method as done by \citet{Kemper04} and \citet{Spoon07}, we measured the ratio of observed flux (f$_{\rm obs}^{\lambda}$) to continuum flux (f$_{cont}^{\lambda}$) at the central wavelength of each silicate feature [S$_{\rm Sil}^{\lambda}=$ln(f$_{\rm obs}^{\lambda}/$f$_{\rm cont}^{\lambda}$)]. For the continuum curve (after removing broad PAH features and narrow emission lines), we assumed a spline (cubic polynomial) with anchor points at 4.8, 5.5, 7.8, 15.0 and 25.0\,$\mu$m {see Fig. \ref{gatos_baseline} in of Appendix \ref{reduction}). The optical depth curve ($\tau_\lambda$) shown in Fig. \ref{gatos_opticaldepth} has been estimated using the same approach as for the silicate bands [i.e. $\tau_\lambda$=-ln(f$_{\rm obs}^{\lambda}/$f$_{\rm cont}^{\lambda}$)] but using every monochromatic wavelength ($\sim$5-30\,$\mu$m). Throughout this work we adopt that S$_{\rm Sil}^{\lambda}$=-$\tau_\lambda$.

All the galaxies studied here show the 9.7\,$\mu$m silicate band in absorption (see Fig. \ref{gatos_opticaldepth} and Table \ref{table_prop}). However, the 18\,$\mu$m feature is relatively weak for the majority of the targets. The only exceptions are NGC\,5728 (S$_{\rm Sil}^{\rm 18\mu m}$=-0.23) and NGC\,7172 (S$_{\rm Sil}^{\rm 18\mu m}$=-0.54). While the spatial scales are similar at 6 and 9.7\,$\mu$m ($\sim$0.3\arcsec), at 18\,$\mu$m it is $\sim$0.7\arcsec. This might slightly affect the 18\,$\mu$m silicate features including extra contribution from the host galaxy. We find the deepest 9.7\,$\mu$m silicate absorption in NGC\,7172 (see also \citealt{Roche07}), NGC\,5728, ESO\,137$-$G034 and NGC\,5506 (S$_{\rm Sil}^{\rm 9.7\mu m}$ of -2.4, -1.9, -1.0, -0.9, respectively). Two of these galaxies (NGC\,5728 and ESO\,137--G034) have high hydrogen column densities (CT-like source), as measured from the X-rays. On the other hand, NGC\,7172 has a modest value of N$_{\rm H}$ from X-rays (\citealt{Ricci17}) and ALMA observations (\citealt{Alonso-Herrero23}) relative to the deep silicate bands found in its nuclear region. However, this galaxy has a dust lane (\citealt{Sharples84}, see also Fig. 1 of \citealt{Alonso-Herrero23}) and, thus, we expect a significant contribution to the silicate strength from this foreground material. Indeed, the contribution of the foreground extinction can be the dominant one in the measured silicate strength of edge-on galaxies (e.g. \citealt{Goulding12,Gonzalez-Martin13}). As shown by high angular resolution ground-based mid-IR spectroscopy of local AGN, the silicate strength alone is not necessarily a good indicator of nuclear obscuration (e.g. \citealt{Alonso-Herrero11,Gonzalez-Martin13,Garcia-Bernete19,Garcia-Bernete22c}).

\begin{figure*}
\centering
\par{
\includegraphics[width=8.3cm]{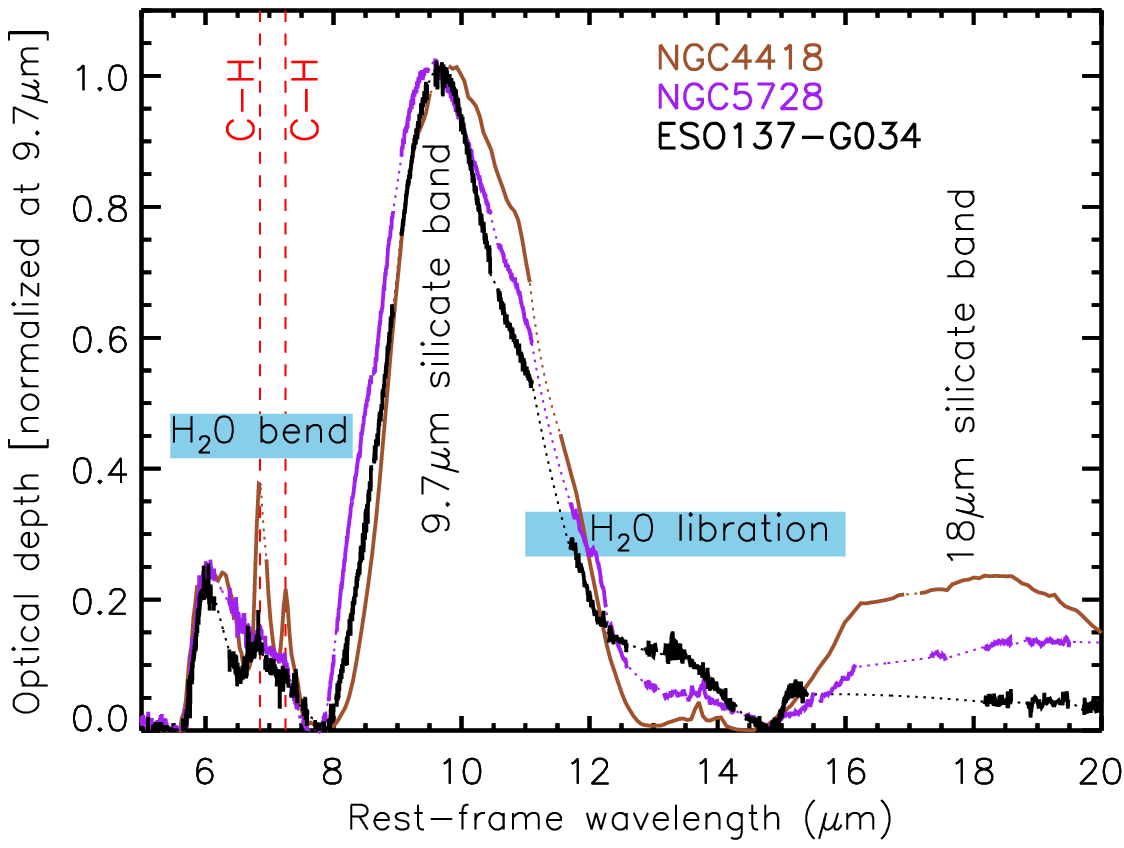}
\includegraphics[width=8.3cm]{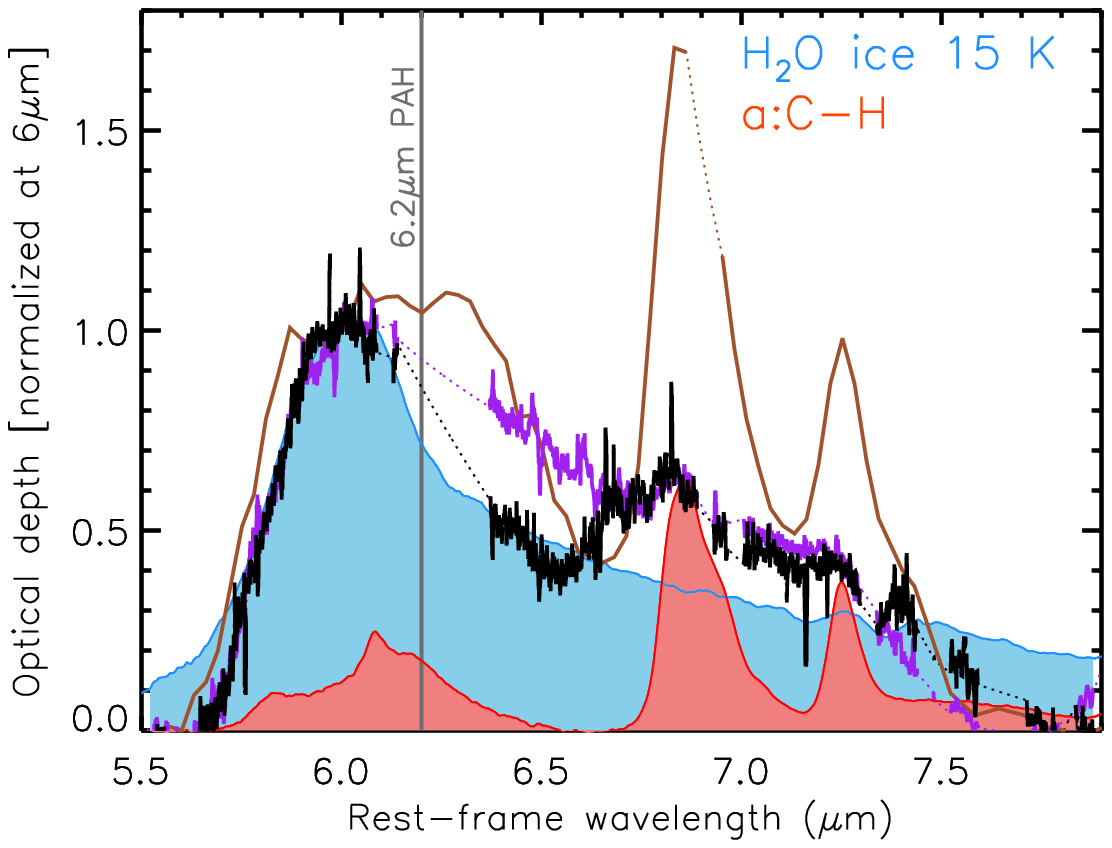}
\par}
\caption{Optical depth of the Compton thick sources normalized at 9.7\,$\mu$m. Left: Observed profiles of the optical depth profiles of NGC\,4418 (brown line), NGC\,5728 (purple line) and ESO\,137-G034 (black solid line). Right: zoom-in of the H$_2$O (bend) absorption band. Laboratory spectra of pure water (blue shaded region corresponds to H$_2$O at 15\,K; \citealt{Ehrenfreund97,Oberg07}) and a a:C–H hydrogenated amorphous carbon analog (red shaded region; \citealt{Dartois07}, see also \citealt{Mate19}) are shown. See Appendix \ref{iceprofile} for further details on the laboratory spectra.}
\label{gatos_zoom}
\end{figure*}

\begin{figure*}
\centering
\par{
\includegraphics[width=8.3cm]{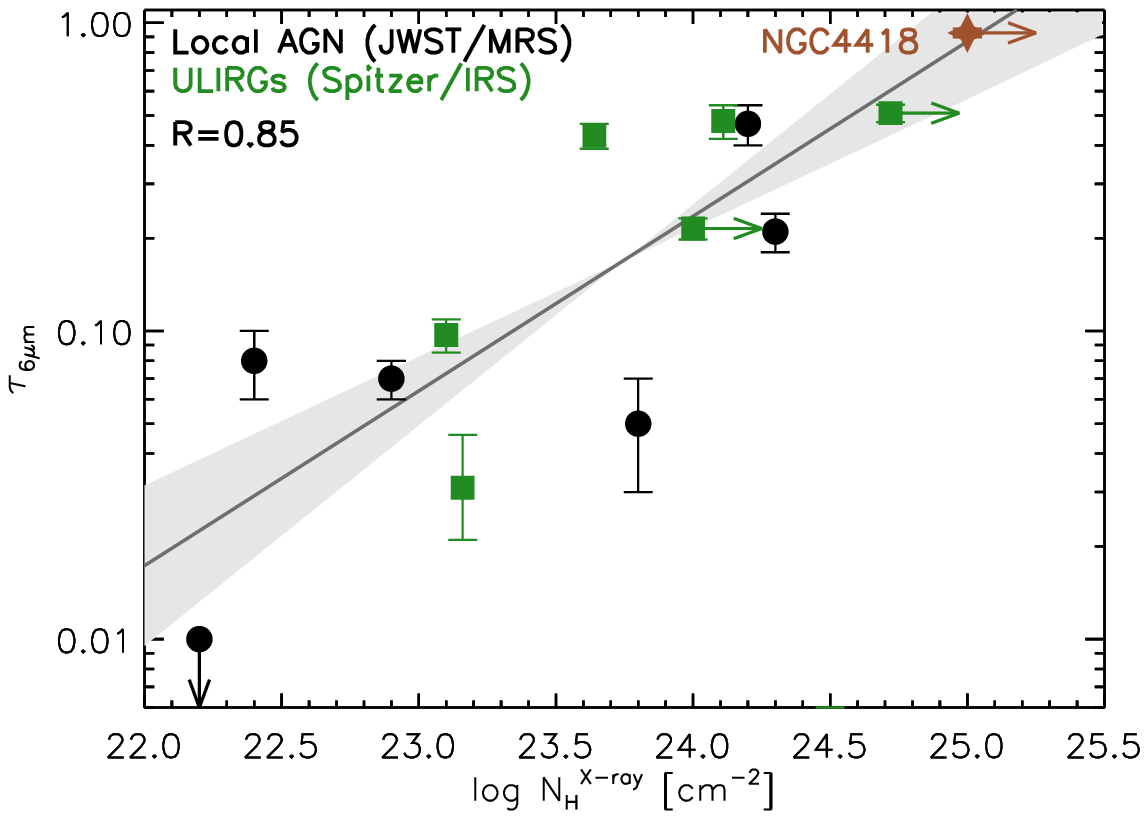}
\includegraphics[width=8.3cm]{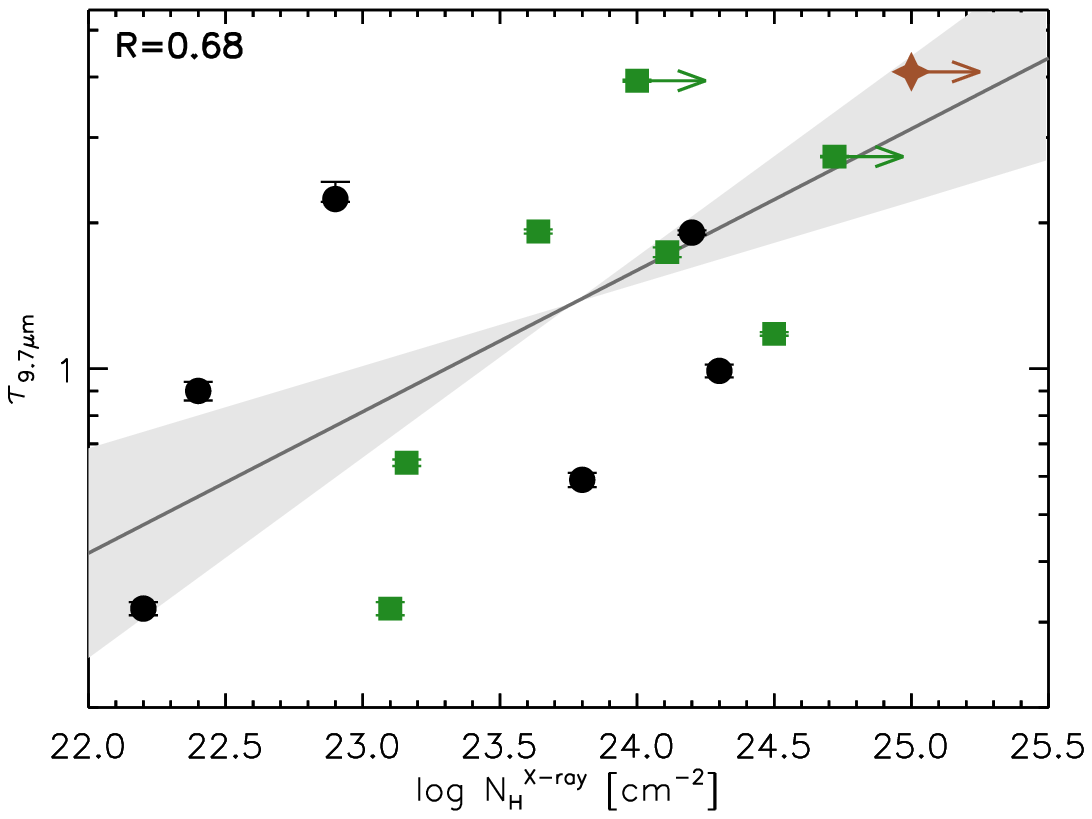}
\par}
\caption{Relationship between X-ray derived column densities and optical depths. We use local obscured AGN and ULIRGs for the correlations presented here. Left: X-ray derived column densities vs. H$_2$O optical depths (R=0.85; log ($\tau_{6\mu m}$)=log (N$_H^{X-ray}$)$\times$ (0.57$\pm$0.15) - 14.24$\pm$3.54). Right: X-ray derived column densities vs. 9.7\,$\mu$m silicate optical depths. Grey dashed lines correspond to the best linear fit (R=0.68; log ($\tau_{6\mu m}$)=log (N$_H^{X-ray}$)$\times$ (0.29$\pm$0.12) - 6.80$\pm$2.90).}
\label{column}
\end{figure*}

\subsection{Water and hydrocarbon ices}
\label{ice}

The nuclear spectra of the sample show extra absorption bands apart from those of the silicates. These features are identified as water (at $\sim$6\,$\mu$m) and aliphatic hydrocarbons ices (at 6.85 and 7.25\,$\mu$m; e.g. \citealt{Spoon22} and references therein).

The 6\,$\mu$m feature, which is associated with {\textit{dirty water ice}} (i.e., a mix of water and other molecules such as CO and CO$_2$), is detected in five of the six nuclei (see Fig. \ref{gatos_opticaldepth}). This feature is only tentatively detected in the case of MCG-05-23-016, which is the source with the lowest N$_H^{X-ray}$ in the sample and is poor in molecular gas \citealt{Rosario18}. This band is prominent in those galaxies with log N$_H^{X-ray}$ (cm$^{-2}$)$\gtrsim$24.0. In Table \ref{table_prop} we present the optical depth of the H$_2$O ($\tau_{6.0\mu m}$) of the sample. $\tau_{6.0\mu m}$ has been estimated using the same approach as for the silicate bands but at 6.0\,$\mu$m.

The aliphatic hydrocarbon bands at 6.85 and 7.25\,$\mu$m (\citealt{Spoon22}) are likely to be present in our obscured AGN, but these are not as prominent as in NGC\,4418 (see Fig. \ref{gatos_zoom}). NGC\,5728 and ESO\,137-G034 also show a shoulder in the 9.7\,$\mu$m silicate absorption band that might be related with the H$_2$O libration band ($\sim$11-16\,$\mu$m). In Fig. \ref{lab} of Appendix \ref{iceprofile}, we show the pure H$_2$O, H$_2$O:CO$_2$ and H$_2$O:CO optical depths measured in laboratory experiments. Given that the sublimation temperature of the CO ices is $\sim$20\,K (e.g. \citealt{Ferrero20,Perrero23}). The comparison suggests that water ice composition is complex and probably might include CO at relatively low temperature ($\sim$20\,K; to decrease the H$_2$O libration mode). This is also compatible with large amount of cold molecular gas present in the nuclear region of AGN (e.g. \citealt{Garcia-Burillo16,Garcia-Burillo21}). However, high angular resolution data at $\sim$4-5\,$\mu$m is need to confirm the presence of the frozen CO band (4.67\,$\mu$m).

\section{Nuclear obscuration in AGN}
\label{nuc_sect3}
Broad-band X-ray derived column densities (N$_H^{X-ray}$) are considered good tracers of nuclear obscuration, even for highly absorbed sources. Indeed, a major step forward was attained with {\textit{NuSTAR}} (3–80\,keV) allowing detailed studes of the X-ray emission in CT AGN (e.g. \citealt{Tanimoto22}). However, even hard X-rays selections are missing a significant fraction of the intrinsic emission in highly absorbed type 2 sources with very high covering factor tori (e.g. \citealt{Mateos17,Garcia-Bernete19,Ricci21}). Furthermore, hydrogen column densities derived from 0.5-10\,keV measurements can be affected by host galaxy contribution in low luminosity AGN (e.g. \citealt{Guainazzi05}). High angular resolution ALMA observations also provide a robust estimation of the nuclear molecular gas column densities (e.g. \citealt{Alonso-Herrero18,Alonso-Herrero23,Combes19,Garcia-Burillo21}), although this might depend on the CO--to--H$_{2}$ conversion factor used. The combination of high angular resolution, sensitivity and wavelength coverage of JWST provides an alternative route for studying the nuclear obscuration in local AGN by using the mid-IR broad absorption bands (i.e. water ices, aliphatic hydrocarbons and silicate features).

\subsection{Water ices as tracers of the innermost obscuration}
\label{tauice}
In this section, we test the reliability of the water ice band as a tracer of the nuclear obscuration. 
First, in Fig. \ref{gatos_zoom} (left) we compare the 6\,$\mu$m water optical depth of the sources with N$_{H}^{X-ray}$ (cm$^{-2}$)>24 where the water ice absorption is prominent (i.e. NGC\,5728, ESO\,137-G034) with that of the extremely obscured nucleus of  luminous infrared galaxy NGC\,4418. We find that the water ice absorption band are similar when normalized to that of the 6\,$\mu$m feature (see right panel of Fig. \ref{gatos_zoom}). The very deep silicate absorption band observed in the nuclear region of NGC\,4418 implies that there is a low contribution of nuclear warm material (\citealt{Roche15}). This indicates that covering factor due to cold material (with T$<$T$_{sub}^{H_2O}$) might be high in obscured AGN.

In Fig. \ref{column}, we compare the H$_2$O and 9.7\,$\mu$m silicate feature optical depths of the sample with the hydrogen column density derived from the X-rays (see Table \ref{table_prop}). To extend the sample, we selected ultraluminous infrared galaxies ($L_{\rm IR}$ (L$_{\odot}$)$>$
10$^{12}$, ULIRGs) from \citet{Ricci21}. We retrieved the 6~$\mu$m ice and 9.7~$\mu$m silicate strengths measured in Spitzer IRS data from the IDEOS database (\citealt{Spoon22}). ULIRGs have very compact mid-IR continuum (e.g. \citealt{Diaz-Santos10}) and, thus, are less affected by contribution from the host galaxy emission to the total mid-IR spectrum as probed by Spitzer. There is a good correlation between $\tau_{6.0\mu m}$ and N$_{H}^{X-ray}$ (R=0.85) even for sources with high hydrogen column densities. The good correlation with N$_H^{X-ray}$ indicates that the water ices are related to the nuclear obscuration, opening an alternative way to measure it in deeply obscured AGN not detected in the X-rays (N$_{H}^{X-ray}$ (cm$^{-2}$)$>$25; e.g. \citealt{Aalto15,Alfonso19}) . In agreement with previous works (\citealt{Goulding12,Gonzalez-Martin13}), the relationship between the 9.7\,$\mu$m silicate feature and N$_{H}^{X-ray}$ shows a larger scatter (R=0.68), indicating that it is not always tracing the nuclear obscuration. 

The silicate strength may be sensitive to the various temperatures in different parts of the torus. The ice feature strength might be proportional to the absorption in the colder parts of the clumps (shielding). In the clumpier structure of the torus of {\textit{less obscured}} AGN (log N$_H^{X-ray}$ (cm$^{-2}$)$\lesssim$23.0), there is a non-zero probability for an AGN-produced photon to escape through the torus along a viewing angle without being absorbed. This can produce a direct view of the broad-line region, regardless of the torus orientation. In regions heated by high-energy photons, from the central engine, ices are unlikely to survive. Exposing ices to the harsh central mechanism radiation field can result in the sublimation of ice grains ($\sim$110-120\,K; \citealt{Fraser01}), but they might survive on the sides of the clumps that are not facing the central source. On the other hand, in high covering factor structures (i.e. almost smooth dust distributions), the ices can be protected by the large amount of dust and gas. Therefore, we suggest that water ice bands might be an alternative and/or complementary tool for tracing the obscuration related to the innermost region of AGN.

\begin{figure}
\centering
\par{
\includegraphics[width=7.0cm]{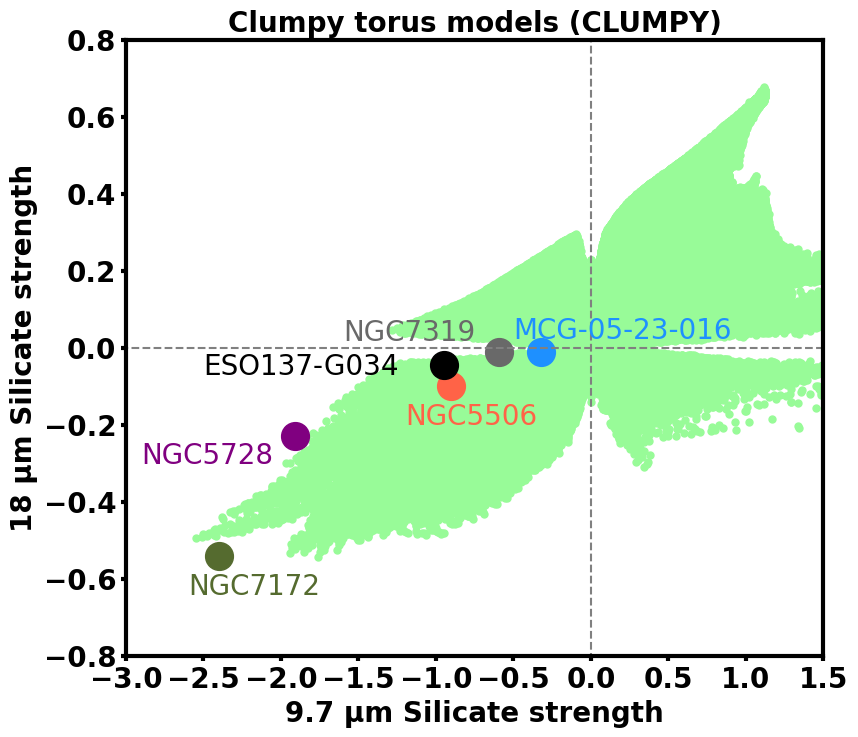}
\includegraphics[width=7.0cm]{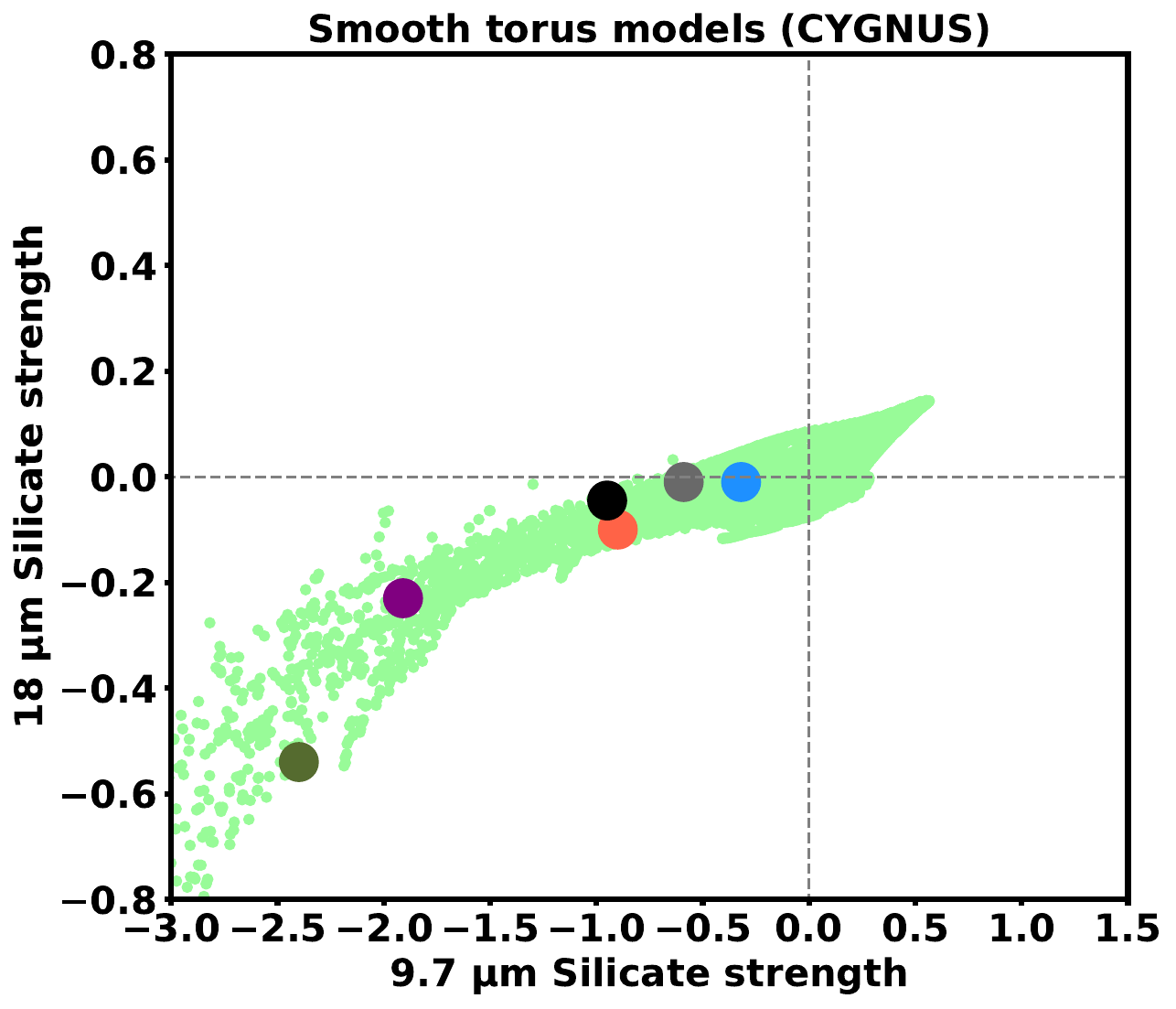}
\par}
\caption{Comparison of the observed 9.7 and 18\,$\mu$m silicate strengths of the sample (big circles of different colors, as in Figs. \ref{gatos} and \ref{gatos_opticaldepth}) with those covered by three different sets of torus models (green dots). Top panel: Clumpy torus models (\citealt{Nenkova08A,Nenkova08B}). Bottom panel: Smooth torus models (\citealt{Efstathiou95,Efstathiou21}). \citet{Martinez-Paredes20} (their Fig. 6) show a similar plot but using additional clumpy and smooth torus models available in the literature.}
\label{nenkova}
\end{figure}

\begin{figure}
\centering
\par{
\includegraphics[width=9.3cm]{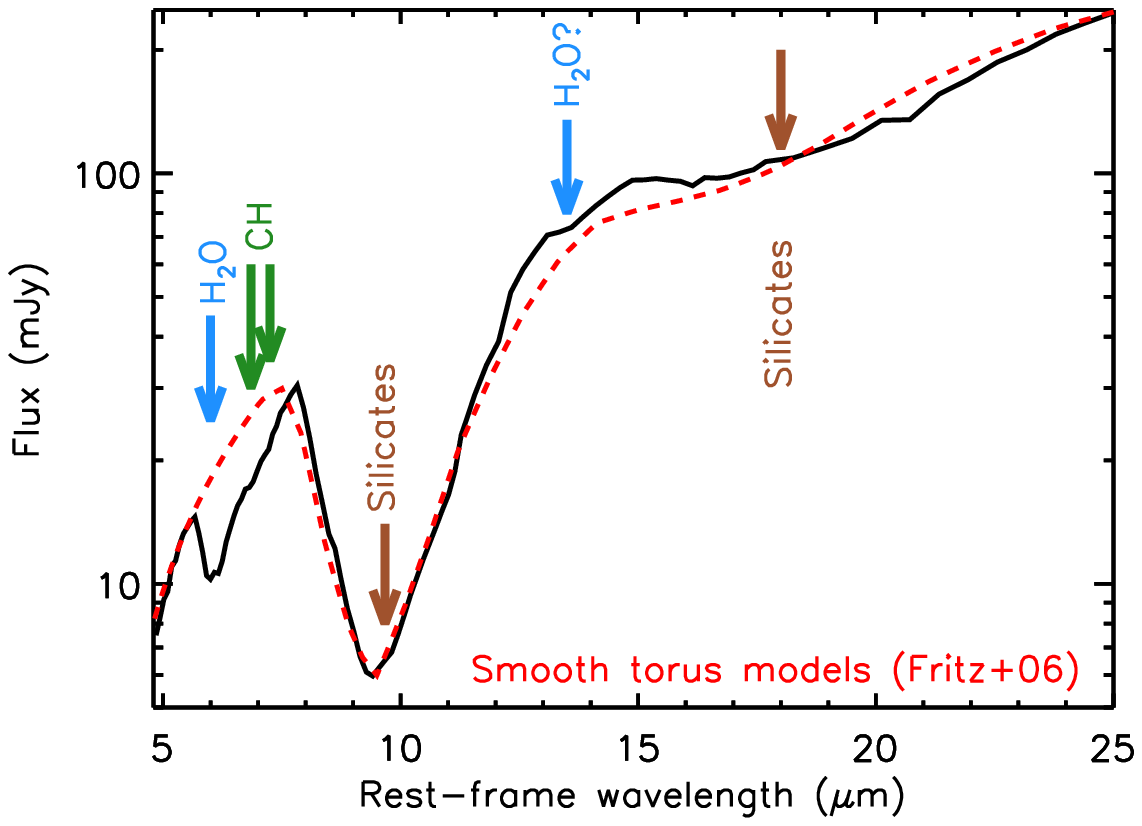}
\par}
\caption{Comparison of the nuclear spectrum of NGC\,5728 (narrow emission lines and PAH features have been removed; red solid line) and the best-fitted torus model (dashed green line) in the 5-25\,$\mu$m region. Blue arrows correspond to the absoption bands of the H$_2$O bend ($\sim$6-8\,$\mu$m) and libration ($\sim$11-16\,$\mu$m) mode. The brown lines represent the 9.7 and 18.0\,$\mu$m silicate features. Green arrows correspond to the absorption bands of the C-H at 6.85 and 7.25\,$\mu$m.}
\label{smooth_torus_short}
\end{figure}

\begin{figure}
\centering
\par{
\includegraphics[width=6.5cm]{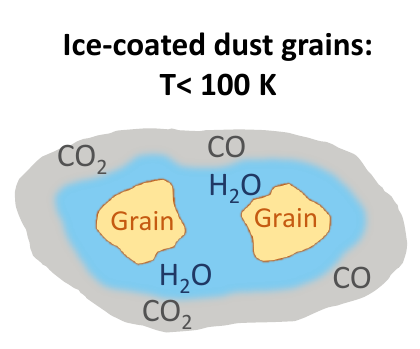}
\par}
\caption{Dust grain sketch showing the expected ice-coated (e.g. H$_2$O, CO, CO$_2$) grains in the outer layers of a nuclear dusty structure with a high covering factor. This is based in the sublimation temperature of the various molecules. Orange, blue and grey colors correspond to naked, H$_2$O and carbon oxides (CO and CO$_2$) ices.}
\label{sketch}
\end{figure}

\subsection{Limitations on the dust composition of current AGN dust models}
\label{limitations}

State-of-the-art torus model have explored different dust distribution (smooth vs. clumpy; \citealt{Fritz06,Nenkova08A,Nenkova08B,Efstathiou95,Efstathiou21}), dust geometries (torus vs. polar dust; e.g. \citealt{Hoenig17,Stalevski19}), dust composition (silicates vs. graphites; e.g. \citealt{Hoenig17,Gonzalez-Martin23}) and various grain sizes (\citealt{Gonzalez-Martin23}).

The relative strengths of the 9.7 and 18\,$\mu$m silicate bands are sensitive to the dust properties and distribution (\citealt{Sirocky08,Thompson09,Tsuchikawa21}). The relationship between these two silicate features and torus model has been previously studied (e.g. \citealt{Sirocky08,Feltre12,Hatziminaoglou15,Martinez-Paredes20,Gonzalez-Martin23}). In Fig. \ref{nenkova}, we compare the observed silicate strengths with two of the torus models available in the literature: clumpy torus (\citealt{Nenkova08A,Nenkova08B}) and smooth torus models (\citealt{Efstathiou95,Efstathiou21}). We find that clumpy and smooth torus models generally explain the observed 9.7 vs. 18\,$\mu$m silicate features. However, the depths of both silicate features for the most obscured AGN in the sample (log N$_H^{X-ray}$ (cm$^{-2}$)$\gtrsim$23.0; i.e. half of the sample) appear to be better explained with models using smooth dust distributions (see Fig. \ref{nenkova}). However, we cannot rule out that extra contribution from the host galaxy is affecting this tentative result, especially for NGC\,7172. We note that fitting the nuclear near and mid-infrared emission with torus models instead of only comparing the silicate feature strengths is necessary to investigate the best suited models representing the nuclear dust emission. This is beyond the scope of this {\textit{letter}}, and will be discussed in a forthcoming paper.

In Fig. \ref{smooth_torus_short} we show an example of SED fitting for nuclear JWST/MRS spectrum of NGC\,5728 (one of the CT AGN in the sample). The smooth torus model by \citet{Fritz06} produces a reasonably good fit to the 9.7\,$\mu$m feature as well as the 5-28\,$\mu$m continuum regions not affected by the absorptions discussed in Section \ref{ice}. However,  Fig. \ref{smooth_torus_short} also suggests the need for developing torus models which include the expected molecular content of the dust mantles (see Fig. \ref{sketch} for a simple sketch representing the expected ice-coated grains in the nuclear dusty structure present in Compton thick AGN and dust-embedded ULIRGs).

\section{Summary and conclusions}
\label{conclusions}
We presented a {\textit{JWST} study of the mid-infrared absorption bands detected in a sample of six obscured type 1.9/2 AGN at distances D$_{L}\sim$30-40\,Mpc. The spatial scales probed by the nuclear spectra of JWST/MRS are $\sim$50~pc at 6\,$\mu$m. The only exception is NGC\,7319, which is part of the Early Release Observations and has a distance of $\sim$100\,Mpc. Finally, our sample of local AGN has different levels of nuclear obscuration ranging from log N$_{\rm H}^{\rm X-Ray}$ (cm$^{-2})\sim22-25$. The main results are as follows. 

\begin{enumerate}
   
\item  We find that all the galaxies show the 9.7\,$\mu$m silicate band in absorption, which is in agreement with their X-ray classification (obscured AGN; i.e. log N$_H^{X-ray}$ (cm$^{-2}$)$>$22.0). However, the 18\,$\mu$m feature is relatively shallow for the majority of the targets.\\

\item  We detect the 6\,$\mu$m {\textit{dirty water ice}} band (i.e. a mix of water and other molecules such as CO and CO$_2$) in five of the six nuclei. This feature is only tentatively detected in the case of MCG-05-23-016, which is the least obscured AGN in the sample (log N$_{\rm H}^{\rm X-Ray}$ (cm$^{-2})\sim22$). Non-detection (or dilution) of water ices in {\textit{less obscured}} AGN might be expected if their tori have lower covering factors and a clumpier medium. Under these conditions, the dusty grains located in the outer layers of the torus, where ices form, can reach hotter temperatures than that of the sublimation of the water ices ($>$T$_{sub}^{H_2O}\sim$110\,K). While this is well documented for many local luminous and ultraluminous infrared galaxies (U/LIRGs), this is the first time that the 6\,$\mu$m is detected in the nuclear region  ($\sim$0.26\arcsec at 6\,$\mu$m; inner $\sim$50\,pc) of local AGN (sub-LIRGs) with different levels of nuclear obscuration ranging from log N$_{\rm H}^{\rm X-Ray}$ (cm$^{-2})\sim22-25$. \\

\item  There is a good correlation between the 6\,$\mu$m water ice optical depths and the hydrogen column densities derived from the X-rays. We include data from ULIRGs with detected water ices to expand the sample. This relationship not only holds for our sample of local AGN but also for higher IR luminosity sources such as ULIRGs ($L_{\rm IR}$ (L$_{\odot}$)$>$
10$^{12}$). In contrast, the relationship between N$_{H}^{x-rays}$ and 9.7\,$\mu$m silicate band strength shows larger scatter, as shown in previous works.\\

\item  By comparing the observed 9.7 and 18\,$\mu$m silicate strengths with dusty torus model predictions, we find that smooth torus models explain better the observed silicate features in the most obscured AGN in our sample (log N$_H^{X-ray}$ (cm$^{-2}$)$\gtrsim$23.0).
   \end{enumerate}

Our results indicate that water ice might be a reliable tracer of the nuclear intrinsic obscuration in AGN. 
The good correlation between the ice optical depth and the X-ray N$_H$ suggests that most of the X-ray absorption occurs in cold and dense clouds. A preliminary comparison with laboratory data suggests that the ice in our local sample of AGN also contains some CO ice at a relatively low temperature ($\sim$20\,K; e.g. \citealt{Ehrenfreund97,Oberg07}). However, to further investigate the composition of ice present in galaxies the whole $\sim$3-15\,$\mu$m range, including the 3 and 6\,$\mu$m H$_2$O ice bands, is needed. We also suggest that it might be necessary to continue exploring the dust composition and grain sizes (e.g. \citealt{Hoenig17,garciagonzalez17,Martinez-Paredes21,Garcia-Bernete22c,Gonzalez-Martin23,Reyes23}) and include the entire molecular content of ice mantles in torus models to better constrain key parameters such as the torus covering factor (i.e. nuclear obscuration).\\

\begin{acknowledgements}

The authors thank E. Dartois for useful discussions, and H.W.W. Spoon for providing laboratory spectra. IGB and DR acknowledge support from STFC through grant ST/S000488/1 and ST/W000903/1. AAH acknowledges support from grant PID2021-124665NB-I00 funded by MCIN/AEI/10.13039/501100011033 and by ERDF A way of making Europe. MPS acknowledges funding support from the Ram\'on y Cajal programme of the Spanish Ministerio de Ciencia e Innovaci\'on (RYC2021-033094-I). OGM acknowledges PAPIIT UNAM IN109123 and to the Ciencia de Frontera project CF-2023-G-100 from CONHACYT. CRA acknowledges the projects ``Feeding and feedback in active galaxies'', with reference PID2019-106027GB-C42, funded by MICINN-AEI/10.13039/501100011033 and ``Quantifying the impact of quasar feedback on galaxy evolution'', with reference EUR2020-112266, funded by MICINN-AEI/10.13039/501100011033 and the European Union NextGenerationEU/PRTR. EB acknowledges the Mar\'ia Zambrano program of the Spanish Ministerio de Universidades funded by the Next Generation European Union and is also partly supported by grant RTI2018-096188-B-I00 funded by the Spanish Ministry of Science and Innovation/State Agency of Research MCIN/AEI/10.13039/501100011033. SGB acknowledges support from the research project PID2019-106027GA-C44 of the Spanish Ministerio de Ciencia e Innovaci\'on. EG-A thanks the Spanish MICINN for support under projects PID2019-105552RB-C41 and PID2022-137779OB-C41. CR acknowledges support from Fondecyt Regular grant 1230345 and ANID BASAL project FB210003. MJW acknowledges the award of a Leverhulme Emeritus Fellowship.

This work is based on observations made with the JWST. The authors acknowledge the ERO team for developing their observing program with a zero--exclusive--access period. The authors are extremely grateful to the JWST helpdesk for their constant and enthusiastic support. Finally, we thank the anonymous referee for their useful comments.
\end{acknowledgements}


\begin{appendix}

\section{{\it JWST} data reduction}
\label{reduction}

We used mid-IR (4.9-28.1 $\mu$m) MIRI MRS integral-field spectroscopy data. The MRS has a spectral resolution of R$\sim$3700--1300 (\citealt{Labiano21}) and comprises four wavelength channels: ch1 (4.9--7.65~$\mu$m), ch2 (7.51--11.71~$\mu$m), ch3 (11.55--18.02~$\mu$m), and ch4 (17.71--28.1~$\mu$m). These channels are further subdivided into three sub-bands (short, medium, and long). The FoV is larger for longer wavelengths: ch1 (3.2\arcsec $\times$ 3.7\arcsec), ch2 (4.0\arcsec $\times$ 4.7\arcsec), ch3 (5.2\arcsec $\times$ 6.1\arcsec), and ch4 (6.6\arcsec $\times$ 7.6\arcsec). We refer the reader to \citet{Argyriou23} and \citet{Rigby23} for further details.

We primarily followed the standard MRS pipeline procedure (e.g. \citealt{Labiano16} and references therein) to reduce the data using the pipeline release 1.11.0 and the calibration context 1095.
Some hot\slash cold pixels are not identified by the current pipeline version, so we added an extra step before creating the data cubes to mask them. We used the background frames to determine the mean flux of the pixels for each channel. Then, we masked those whose flux deviates more than $\pm$6$\sigma$. After masking these bad pixels, we tried to recover their true values by linear interpolation of the fluxes of the pixels above and below in the same column (i.e., approximate spectral axis). We limited this interpolation to pixels where the signal-to-noise ratio (SNR) was $>$30. Based on our tests, this interpolation approach is able to recover the actual flux within $\pm$10\% for this SNR threshold. 
To subtract the background, we first created background-only cubes for each sub-band of each channel using the dedicated background observations. Then, we generated the background spectra by averaging these cubes in each spectral channel and these spectra were subtracted from the science cubes.

The nuclear spectra from the different sub-channels were extracted assuming they are point sources. Using the cube oriented in the instrument integral field unit plane, we employed observations of calibration point sources (MRS HD-163466 and IRAS,05248$-$7007, Program IDs 1050 and 1049) to measure the width and position angle of a 2D Gaussian for each spectral channel. To obtain the point source flux we used the models of the calibration PSF stars from \citet{Bohlin20}, which is equivalent to applying aperture correction factors. This 2D Gaussian represents the point-spread function of MRS, whose width increases from shorter to longer wavelengths. Subsequently, we fitted the amplitude and position of the 2D Gaussian to
the nuclear regions of our AGN sample, while keeping fixed the width and position angle values obtained from the calibration point source. Fig. \ref{gatos_baseline} present the JWST/MRS spectra (PAH features and narrow emission lines masked) and the fitted continuum baseline (see Section \ref{nuc_sect3}).

\begin{figure*}
\centering
\par{
\includegraphics[width=15.3cm]{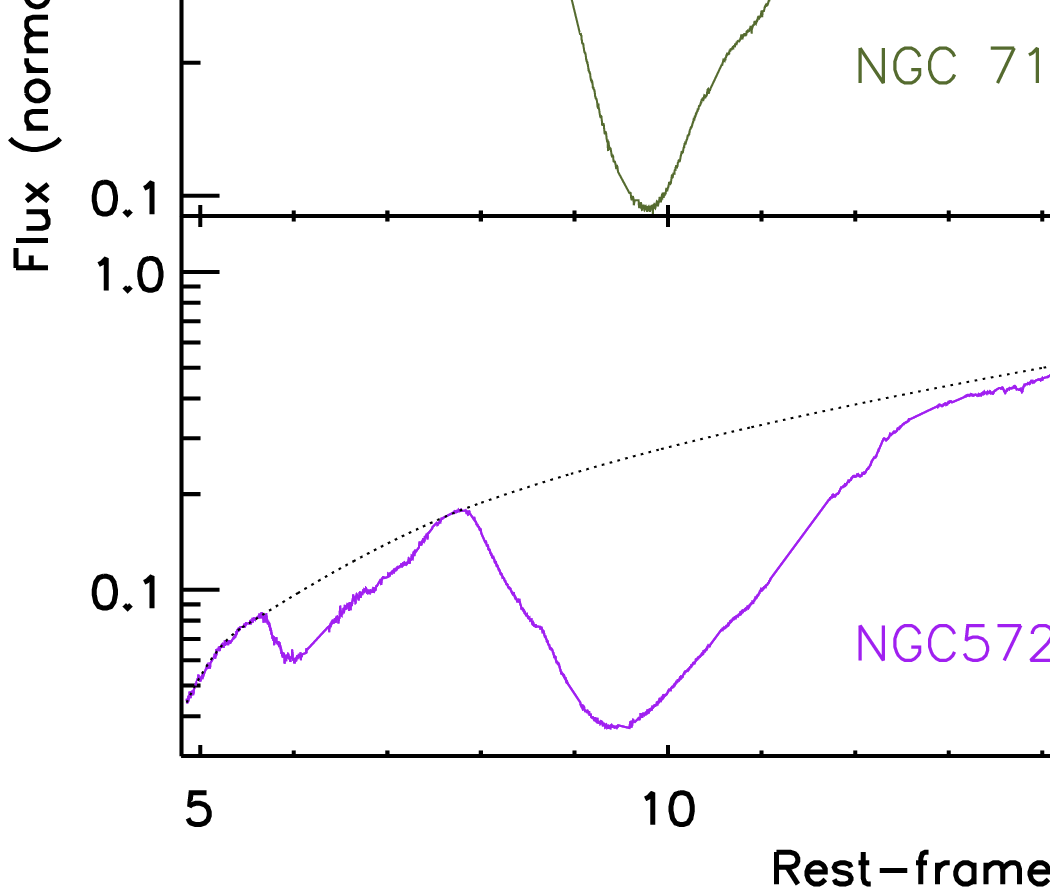}
\par}
\caption{JWST/MRS nuclear spectra of our objects (PAH features and narrow emission lines masked). The continuum baseline is shown in black (dotted line).}
\label{gatos_baseline}
\end{figure*}

\section{Archival {\it Spitzer}/IRS data of NGC\,4418}
\label{ngc4418}

We retrieved low-resolution (R$\sim$60--120) mid-IR staring spectrum of NGC\,4418 from the Cornell Atlas of Spitzer/IRS Source (CASSIS, version LR7; \citealt{Lebouteiller11}). The spectra were reduced with the CASSIS software, using the optimal extraction to get the best signal-to-noise ratio, which is equivalent to a point source extraction (see Fig. \ref{ngc4418_fig}). We note that \citet{Spoon22} measured for this galaxy $\tau_{6\,\mu m}$ and $\tau_{9.7\,\mu m}$ of 0.93$\pm$0.02 and 4.12$\pm$0.02, respectively. These are in good agreement, within the errors, with the values reported in Table \ref{table_prop}.

\begin{figure}
\centering
\par{
\includegraphics[width=10.3cm]{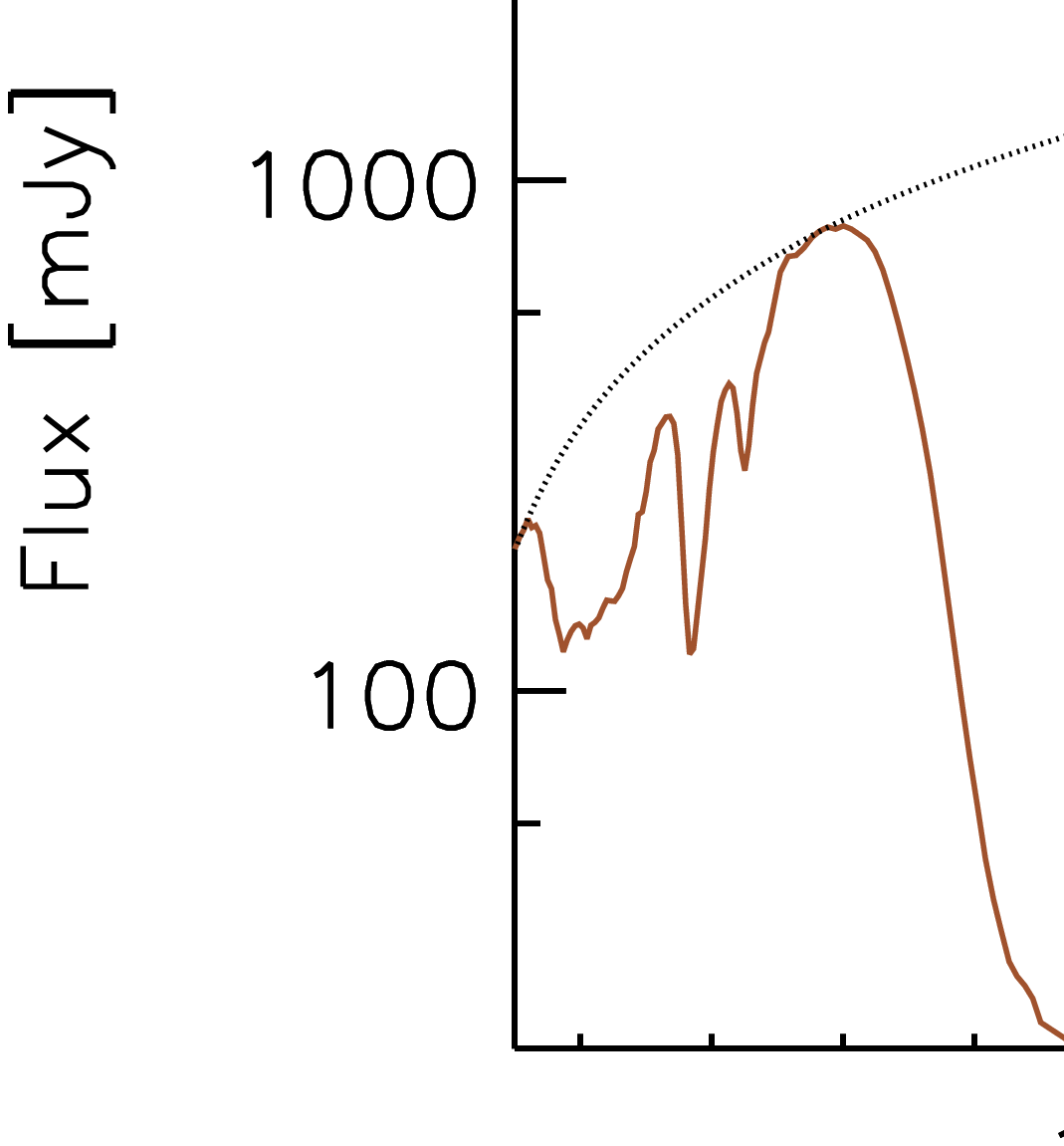}
\par}
\caption{Spitzer/IRS nuclear spectrum of NGC\,4418 (PAH features and narrow emission lines masked). The continuum baseline is shown in black (dotted line).}
\label{ngc4418_fig}
\end{figure}

\section{H$_{2}$O ice profile}
\label{iceprofile}
In this Section we present the laboratory spectra of pure water (see Fig. \ref{lab} and \ref{lab2}). The data presented in this section was downloaded from the Leiden Ice Database (Rocha et al. in prep). Fig. \ref{gatos_zoom_lab} shows a comparison between the 6\,$\mu$m absorption band of NGC\,4418, NGC\,5728 and ESO\,137-G004 and laboratory data.

\begin{figure*}
\centering
\par{
\includegraphics[width=13.3cm]{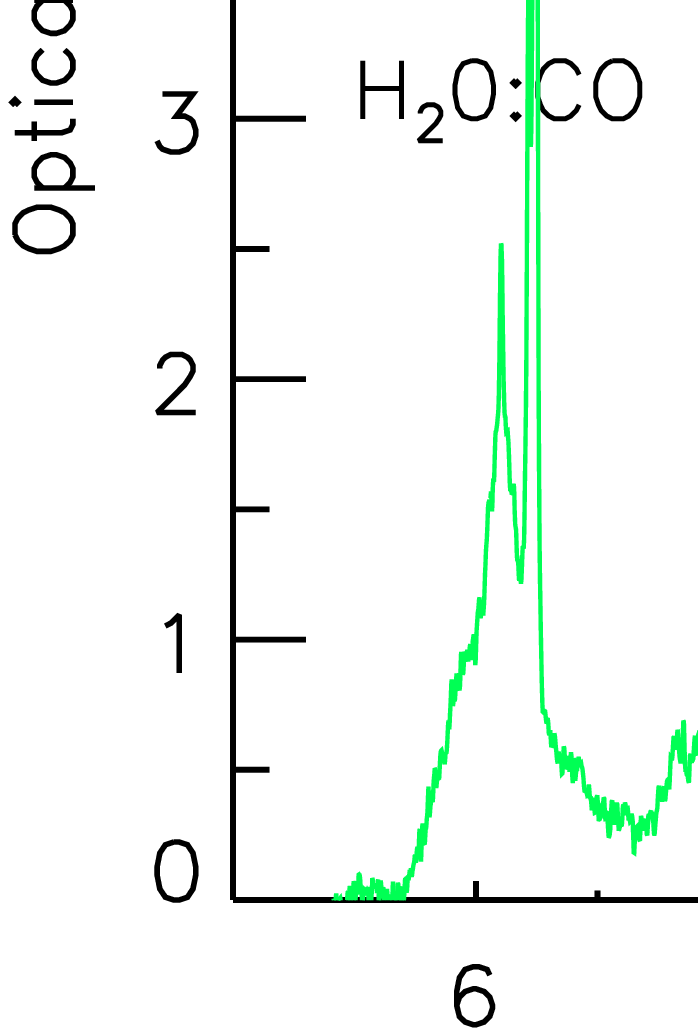}
\par}
\caption{Laboratory measurements of the water ices optical depth as a function of temperature. Top panel: pure water. Central panel: water and CO$_{2}$ mix (1:4). Bottom panel: water CO mix (1:100). Laboratory data are from \citet{Ehrenfreund97} and \citet{Oberg07}. See Fig. 29 of \citealt{Spoon22} for a similar plot.}
\label{lab}
\end{figure*}

\begin{figure*}
\centering
\par{
\includegraphics[width=13.3cm]{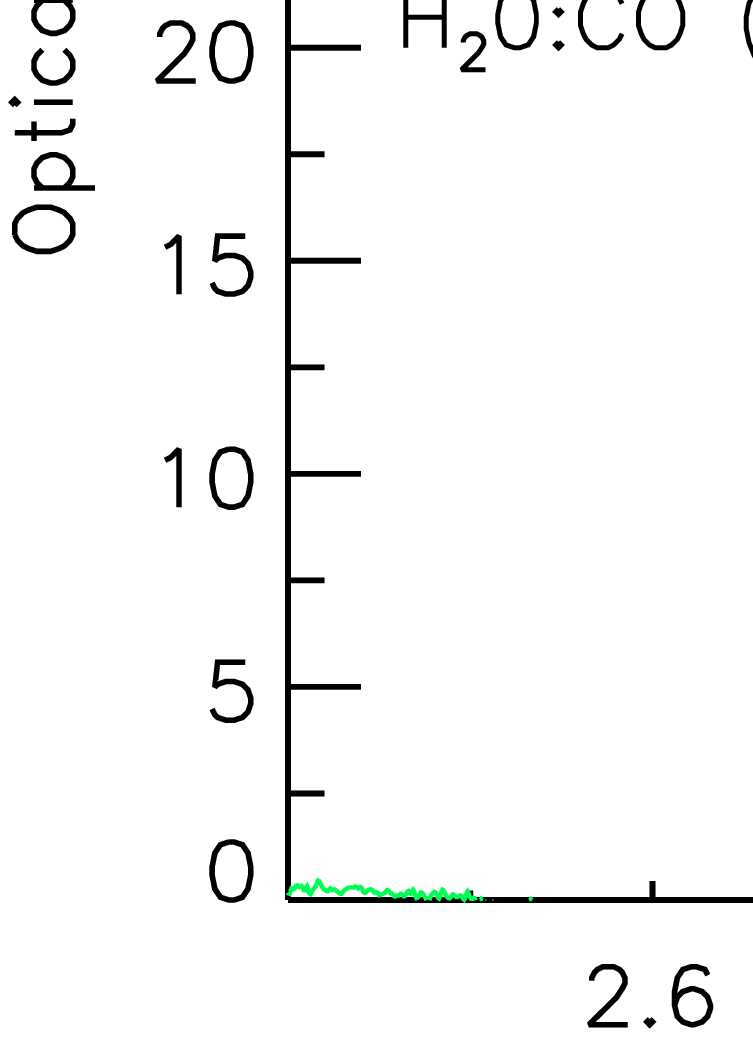}
\par}
\caption{Same as Fig. \ref{lab} but showing the stretch mode ($\sim$3\,$\mu$m) of the water ice band.}
\label{lab2}
\end{figure*}

\begin{figure}
\centering
\par{
\includegraphics[width=8.3cm]{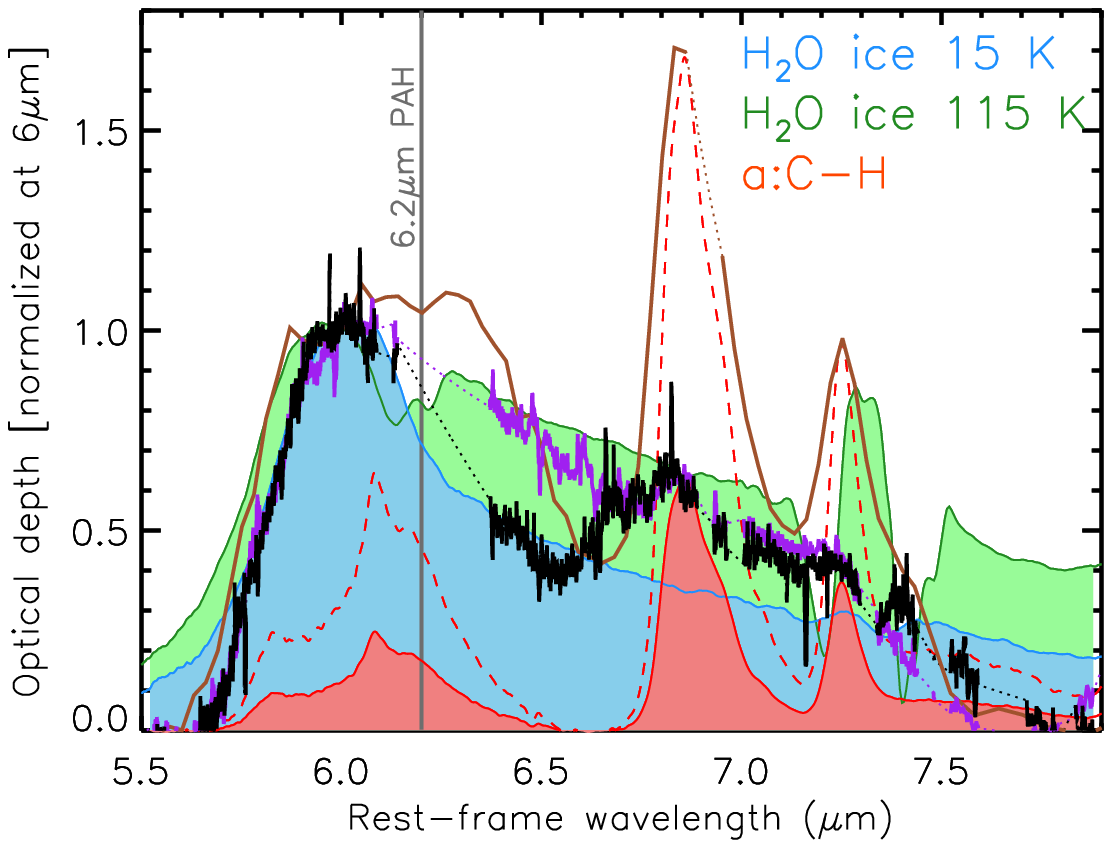}
\par}
\caption{Optical depth of the Compton thick sources normalized at 9.7\,$\mu$m. Left: optical depth profiles of NGC\,4418 (brown line), NGC\,5728 (purple line) and ESO\,137-G034 (black solid line). Right: zoom-in of the H$_2$O (bend) absorption band. Laboratory spectra of pure water (blue and green shaded regions correspond to H$_2$O at 15 and 115\,K, respectively; \citealt{Ehrenfreund97,Oberg07}) and a a:C–H hydrogenated amorphous carbon analog (red shaded region; \citealt{Dartois07}, see also \citealt{Mate19}) are shown. See \citet{Spoon22} for a similar plot (their Fig. 29).}
\label{gatos_zoom_lab}
\end{figure}

\end{appendix}

\end{document}